\documentclass[aps,prb,twocolumn,groupedaddress]{revtex4}

\usepackage{amsmath}
\usepackage{graphics}
\usepackage{graphicx}
\usepackage{amsmath}
\usepackage{amsfonts}
\usepackage{amssymb}

\def \beq{\begin{equation}}
\def \eeq{\end{equation}}
\def \beqar{\begin{eqnarray}}
\def \eeqar{\end{eqnarray}}

\begin{document}

\title{\bf{Intergranular pinning potential and critical current in the magnetic
superconductor RuSr$_{2}$Gd$_{1.5}$Ce$_{0.5}$Cu$_{2}$O$_{10}$}}

\author{M.G. das Virgens$^{1,2}$, S. Garc\'{\i}a$^{1,3}$, M.A. Continentino$^2$ and L. Ghivelder$^1$}

\affiliation{$^1$Instituto de F\'{\i}sica, Universidade Federal do
Rio de Janeiro, C.P. 68528, Rio de Janeiro, RJ 21941-972,
Brazil\\$^2$Instituto de F\'{\i}sica, Universidade Federal Fluminense, C.P. 68528, Niter%
\'{o}i, RJ 21945-970, Brazil\\$^3$Laboratorio de
Superconductividad, Facultad de F\'{\i}sica-IMRE, Universidad de
La Habana, San L\'{a}zaro y L, Ciudad de La Habana 10400, Cuba}


\begin{abstract}
The intergranular pinning potential $U$ and the critical current density $%
J_{C}$ for polycrystalline
RuSr$_{2}$Gd$_{1.5}$Ce$_{0.5}$Cu$_{2}$O$_{10}$ ruthenate-cuprate
were determined at zero magnetic field and temperature through the
frequency shift in the peak of the imaginary part of the ac
magnetic susceptibility, $\chi ^{\prime \prime }.$\ A critical
state model, including a flux creep term, was found to accurately
describe the $\chi ^{\prime \prime }$\ behavior. The obtained
values, $U$($H$=0,$T$=0) $\cong $ 30 meV and \
$J_{C}$($H$=0,$T$=0) $\cong $ 110 A/cm$^{2}$ are about two orders
of magnitude and four times lower, respectively, in comparison
with the high-T$_{c}$ cuprate YBa$_{2}$Cu$_{3}$O$_{7}$. These
results were ascribed to the effects of the Ru magnetization on
the connectivity of the
weak-linked network, giving an intrinsic local field at the junctions of $%
\sim $ 15 Oe. The impact on $J_{C}$ is less intense because of the small
average grain radius ($\sim $1 $\mu $m). The intragranular London
penetration length at $T$ = 0, [$\lambda _{L}$(0) $\cong $ 2 $\mu $m], was
derived using a Kim-type expression for the field dependence of $J_{C}$. A
possible source for the large value of $\lambda _{L}$ in comparison to the
high-$T_{c}$ cuprates is suggested to come from a strong intragrain
granularity, due to structural domains of coherent rotated RuO$_{6}$
octahedra separated by antiphase boundaries.
\end{abstract}

\maketitle

\section{Introduction}

The coexistence of ferromagnetic (FM) long-range order of the Ru moments
with a superconducting (SC) state in the ruthenate-cuprates RuSr$_{2}$RCu$%
_{2}$O$_{8}$ (Ru-1212) and RuSr$_{2}$(R,Ce)$_{2}$Cu$_{2}$O$_{10}$ (Ru-1222),
where R = Gd, Eu, has been intensively studied in the recent years.\cite%
{Bauernfeind, McLaughlin, Bernhard, Liu, Felner02, Xue02, Williams02,
Williams03} Among several major open issues, the interplay between the
transport properties and magnetism has received a great deal of attention.%
\cite{Tallon, Croone, Chen01, Pozek02} On the other hand, there are few
reports on the intergrain properties, which exhibit very interesting
features. The broad resistive SC transition ($\Delta T_{SC}$ $\approx $
15-20 K) observed in good quality Ru-1212 ceramic samples has been
consistently explained in terms of a strong intergrain contribution and
spontaneous vortex phase formation in the grains, as evidenced through
microwave resistivity measurements in powders dispersed in an epoxy resin.%
\cite{Pozek01} An abrupt reduction in the suppression rate of the
intergranular flux activation energy with the increase in magnetic field has
been observed at $H$ = 0.1 T in polycrystalline Ru-1212, through a study of
the $I$-$V$ characteristic curves.\cite{Garcia03b} This behavior has been
ascribed to a spin-flop transition of the Ru-sub-lattice, leading to a
decrease of the effective local field at the junctions. In addition, a
preliminary report\cite{Felner03}on $I$-$V$ curves for Ru-1222 suggests that
the low values obtained for the intergranular current density $J_{C}$ is
possibly related to its magnetic behavior, indicating that more
investigation is needed to clarify this point. The peak of the SC intergrain
transition, as determined from the derivative of the resistive curves, is
quite intense and narrow in both ruthenate-cuprate systems.\cite{Garcia03a}
This result is at variance with the high-$T_{c}$ cuprates, exhibiting a
smeared intergrain peak of small amplitude as a consequence of a
broad-in-temperature phase-lock process across a wide distribution of link
qualities. The results for the ruthenate-cuprates have been interpreted as a
consequence of the effects of the Ru-magnetization on the grain boundaries,
in such a way that the intergrain percolation occurs only through a fraction
of high quality junctions.\cite{Garcia03a} These reports clearly evidence
that the magnetization in the grains leaves a sizable effect in the
connectivity of the weak link network. Whether this unique feature of the
ruthenate-cuprates changes the essentials of the intergrain properties in
comparison to the high-$T_{c}$ superconductors remains an open issue; a
quantitative determination of the parameters characterizing the intergrain
coupling in these compounds is still lacking.

Since the early works in the high-$T_{c}$ superconductors, ac
magnetic susceptibility has proved to be a useful tool to
characterize their granular properties.\cite{Goldfarb, Gomory,
Chen88} In particular, a critical state model for granular
superconductors was used to calculate the temperature, and both ac
and dc magnetic field dependence of the complex susceptibility,
$\chi=\chi^{\,\prime}+i\,\chi^{\prime\prime}$,
for sintered bulk samples of YBa$%
_{2}$Cu$_{3}$O$_{7}$ (YBCO),\cite{Muller89, Nikolo94a, Ghiv94} (Bi,Pb)$_{2}$%
Sr$_{2}$Ca$_{2}$Cu$_{3}$O$_{10}$ and Bi$_{2}$Sr$_{2}$CaCu$_{2}$O$_{8}$,\cite%
{Muller91} with excellent results. In this investigation we present detailed
measurements of the ac magnetic susceptibility as a function of frequency $f$
and amplitude $h_{ac}$ of the driving field in polycrystalline RuSr$_{2}$Gd$%
_{1.5}$Ce$_{0.5}$Cu$_{2}$O$_{10-\delta }$. We show that the simple models
proposed to describe the dependence on these parameters of the imaginary
part of the susceptibility, $\chi ^{\prime \prime }$, in the conventional
cuprates\cite{Nikolo89, Muller90, Muller91, Nikolo94b} quantitatively
accounts for the behavior of the dissipation peak in Ru-1222. Nevertheless,
the parameters characterizing the intergrain properties, as the pinning
potential depth, the full penetration field and the critical current
density, are considerably lower in Ru-1222 as compared to the high-$T_{c}$
cuprates. We propose that these results are due to effects of the Ru
magnetization within the grains on the connectivity of the weak link
network. To the best of our knowledge, there are no previous reports which
quantitatively evaluates these magnitudes in Ru-1222.

\section{Experimental}

Polycrystalline RuSr$_{2}$Gd$_{1.5}$Ce$_{0.5}$Cu$_{2}$O$_{10-\delta }$
(Ru-1222) was prepared by conventional solid-state reaction. Details on
sample preparation, microstructure and the resistive SC transition can be
found elsewhere.\cite{Garcia03a} The room temperature x-ray diffraction
pattern corresponds to Ru-1222, with no spurious lines being observed.
Scanning electron microscopy revealed a dense packing of grains, with an
average grain radius $\cong $ 1 $\mu $m, leading to a well-connected
microstructure. The intragrain SC transition temperature is $T_{SC}$ = 22 K.%
\cite{Garcia03a} Bars of $\simeq $ 10 x 1.7 x 1.7 mm$^{3}$ were
cut from the sintered pellet. The real ($\chi ^{\prime }$) and
imaginary ($\chi ^{\prime \prime }$) parts of the ac magnetic
susceptibility were measured in a Quantum Design PPMS system, with
ac amplitudes $h_{ac}$ = 0.03, 0.1, 0.3, and 1 Oe and frequencies
$f$ = 35, 100, 350 Hz, and 1.0, 3.5, and 10 kHz. For $f$ = 1 kHz
the curves with $h_{ac}$ = 0.01 and 3 Oe were also measured. The
$\chi ^{\prime \prime }$ peak temperature, $T_{P}$, was determined
by conventional numerical derivation of the experimental
curves.\cite{Pureur}

\begin{figure}[h]
\centering
\includegraphics[scale=0.55]{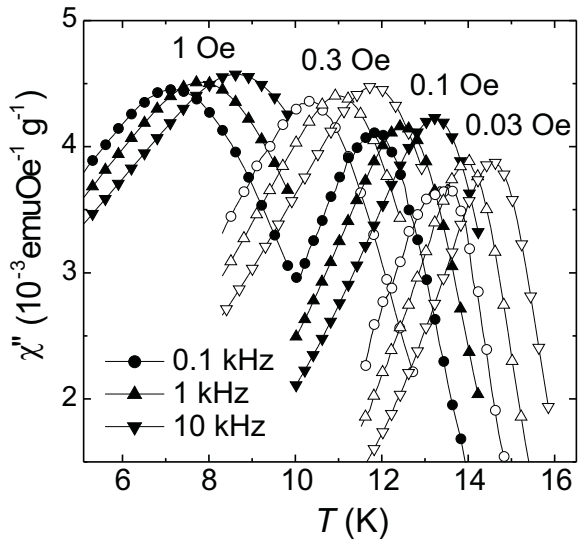}
\caption{Temperature dependence of the imaginary part of the ac
magnetic susceptibility,$\chi ^{\prime \prime }$, for Ru-1222
using different amplitudes and frequencies of the driving field.
Selected curves for $f$ = 0.1, 1, and 10 kHz are shown. For the
sake of clarity not all measured data points are plotted.}
\end{figure}

\section{Results}

Figure 1 shows the imaginary part of the ac susceptibility of Ru-1222, $\chi
^{\prime \prime }$($h_{ac}$,$T$,$f$). For the sake of clarity, only selected
curves are presented, for low, medium and high frequency values. The peaks
in $\chi ^{\prime \prime }$\ are grouped in four sets of data, corresponding
to the ac amplitudes $h_{ac}$ used. Higher $h_{ac}$ values move the curves
to lower temperatures. Inside each set of data, the peaks are shifted to
higher temperatures as $f$ is increased. Typically, the total temperature
shift, $T_{P}$, for the measuring $\Delta f$ interval, is approximately 2 K.
In Fig. 2 we show a logarithmic plot of $f$ vs. $1/T_{P}$ for different $%
h_{ac}$ values. The points clearly follow a linear dependence, corresponding
to an Arrhenius-type expression $f$ = $f_{0}\exp $($E_{f}$/$kT_{P}$), where $%
E_{f}$ plays the role of an activation energy associated to the frequency
effects, $f_{0}$ is a characteristic frequency and $k$ is the Boltzmann
constant. The activation energies calculated from the slopes of the linear
fits of the Arrhenius plots are in the 8-30 meV range. The inset of Fig. 3
shows $E_{f}$ as a function of $h_{ac}$ using a logarithmic scale. For
comparisons with the results obtained for the high-$T_{c}$ cuprates, we
extrapolated to $h_{ac}$ = 0.02 Oe ($\cong $ 1 A/m rms), giving $E_{f}$($%
h_{ac}$=0.02 Oe) $\cong $ 30 meV.

\begin{figure}[h]
\centering
\includegraphics[scale=0.55]{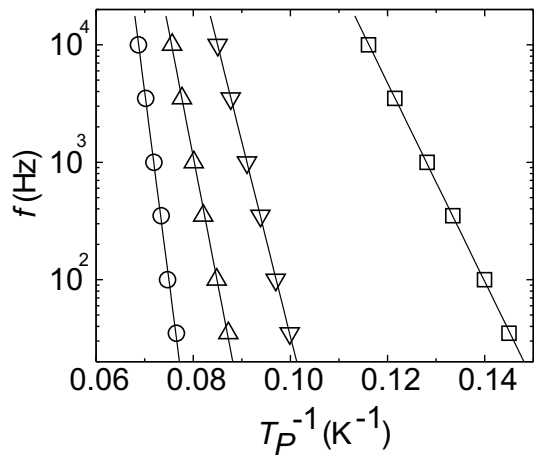}
\caption{Logarithmic plot of the ac driving frequency versus the
reciprocal
of the $\chi ^{\prime \prime }$-peak temperature $T_{P}^{-1}$ for different $%
h_{ac}$ amplitudes; from left to right $h_{ac}$ = 0.03, 0.1, 0.3,
and 1 Oe. The continuous lines are linear fittings.}
\end{figure}

The variation in $T_{P}$ for different $h_{ac}$ amplitudes,
measured with $f$ = 1 kHz, is plotted in Fig. 3. The largest
measuring amplitude for which a peak is observed is $h_{ac}$ = 3
Oe, with $T_{P}$ = 2.7 K. A polynomial fitting (continuous line)
yield $h_{ac}$($T_{P}$=$0$) = 5 Oe. This is the full penetration
field $H$*(0) of the bar-shaped sample at $T$ = $0$ for $f$ = 1
kHz.

\section{Discussion and Conclusions}

As already mentioned, the critical state model has been successfully used to
characterize the granular properties of high-$T_{c}$ superconductors. In
particular, it was shown that a flux creep term added to the current density
term in the critical state equation accurately accounts for the shift in the
$\chi ^{\prime \prime }$ peak to higher temperatures with increasing
frequency of the ac field.\cite{Muller91, Nikolo89, Muller90, Nikolo94b} In
the following, we briefly review the essential ideas of this approach
necessary to conduct the discussion; details can be found in the previously
mentioned reports. We compare our results with bulk YBCO, for which more
detailed data is available. One interesting prediction of the model is that
larger shifts should be observed with decreasing average grain size and
critical current density at the intergrain junctions.\cite{Muller90} Since
the connectivity of the weak link network in Ru-1222 is expected to be
affected by the Ru-magnetization in the grains and the Ru-1222 average grain
radius $R_{g}$ $\approx $ 1 $\mu $m is lower than those reported for the
YBCO sample ($\sim $7-10 $\mu $m), this compound is a suitable material to
verify this point. The five times larger temperature shift $\Delta T_{P}$
due to a frequency variation, observed for Ru-1222 as compared with YBCO [$%
\Delta T_{P}$(YBCO)$\sim $0.4 K \cite{Muller90}] for the same $\Delta f$
interval confirms these predictions. It is worth mentioning that $H$*($0$)
for the Ru-1222 bulk sample is reduced by a factor of four [$\Delta H$*($0$)$%
_{YBCO}$ = 20 Oe for $f$ = 1 kHz \cite{Nikolo94b}], also pointing to a weak
intergrain connectivity. The dense, well-connected microstructure observed
in the SEM images indicate that this poor intergranular coupling is not
related to small contact areas between the grains.

\begin{figure}[h]
\centering
\includegraphics[scale=0.5]{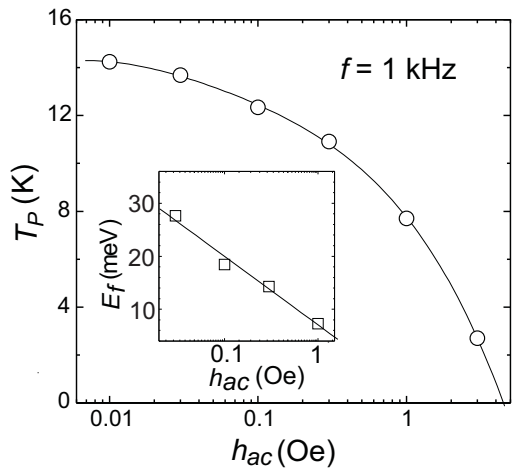}
\caption{The dependence of the $\chi ^{\prime \prime }$-peak temperature $%
T_{P}$ on log $h_{ac}$. The continuous line is a polynomial fitting, giving $%
h_{ac}$ = 5 Oe for $T_{P}$ = 0. Inset: The dependence of the
activation energies $E_{f}$, as determined from the slopes of the
linear fittings in Fig. 2, on $\log h_{ac}$. The extrapolation to
$h_{ac}$ = 0.02 Oe ($\sim $ 1 A/m rms) gives $E_{f}$($h_{ac}$=0.02
 Oe)$\,\approx$ 30 meV.}
\end{figure}

The shift in $T_{P}$ can be understood in terms of a frequency dependent
effective pinning. As the frequency is increased, the intergranular vortices
have less time to creep into the superconductor during each ac cycle. Then,
in order to obtain the full penetration condition, a weaker intergranular
pinning force density is needed to compensate a less efficient creep. Since
the pinning force density weakens with increasing temperature, $T_{P}$ must
increase with the rise in frequency.

For the calculation of the intergranular critical current density
the critical state equation for an infinite slab is
used:\cite{Clem, De Gennes}

\begin{equation}
dH(x)/dx=J_{C}(H,T)=\frac{F_{P}(H,T)}{\left\vert B(x)\right\vert }
\end{equation}%
\newline
where $B$($x$) is the local macroscopic flux density due to the external
field, $x$ is the coordinate perpendicular to the slab, and $F_{P}$($H$,$T$)
is the pinning force density given by

\begin{equation}
F_{P}(H,T)=\frac{1}{V_{b}~d}~\left[ U(H,T)+kT\ln \left( \frac{f}{f_{0}}%
\right) \right]
\end{equation}%
\newline
where $U$($H,T$) is the pinning depth potential, $V_{b}$ is the flux bundle
volume and $d$ half the width of the pinning potential well. The second term
in Eq. (2) represents the flux creep contribution. The flux bundle $V_{b}$
for intergranular vortices is assumed to contain a single flux quantum $%
\mathit{\Phi }_{0}$, and the pinning sites are supposed to be located
between the corners of adjacent grains. For granular material approximated
by a regular array of junctions and cubic grains it is obtained\cite%
{Muller90}

\begin{equation}
V_{b}=\frac{2R_{g}\mathit{\Phi }_{0}}{\left\vert B(x)\right\vert }
\end{equation}%
\newline
where $R_{g}$ represents the average grain radius. The half width $d$ of the
pinning potential is equated to $R_{g}$. Inserting Eqs. (2) and (3) into Eq.
(1) and evaluating for$\ H$=0 and $T$=0 one obtains

\begin{equation}
J_{C}(0,0)_{Ru}=\frac{U(0,0)}{2R_{g}^{2}~\mathit{\Phi }_{0}}
\end{equation}%
\newline
where $J_{C}$(0, 0)$_{Ru}$ is the critical current density at zero \textit{%
external} field of a weak link network with its connectivity affected by the
intrinsic Ru-magnetization; we will return to this point below.

M\"{u}ller\cite{Muller90} showed that in the zero-field limit $E_{f}$($T_{P}$%
, $h_{ac}\cong $0)$~\cong U$(0,0). The procedure adopted in the studies of
high-$T_{c}$ cuprates is to extrapolate the $E_{f}$ vs. $\log h_{ac}$ plot
to the low-field region. Since the extrapolation to exactly $h_{ac}$=0 is
not possible, due to the divergence of the logarithmic function, the usual
practice is to take $h_{ac}$=1 A/m (rms) $\cong $ 0.02 Oe as a reference
\textquotedblleft low field criterion\textquotedblright ,\cite%
{Muller91,Nikolo89,Muller90} which is close to our lowest $h_{ac}$ amplitude
used. The extrapolation in Fig. 3 gives $U$(0, 0)$\cong $ 30 meV. This value
represents a 400-fold decrease in comparison to YBCO [$U$(0,0)$_{YBCO}\cong ~
$12 eV]. Using Eq. (4) with $U$(0, 0)$~\cong ~$30 meV we obtain $J_{C}$(0, 0)%
$_{Ru}$ = 110 A/cm$^{2}$, a value six times lower in comparison to YBCO.\cite%
{Nikolo94b} It is important to notice that the decreasing factor for $J_{C}$%
(0,0), as determined from the frequency shift, is similar to that for $H$%
*(0), which was obtained from the $h_{ac}$ dependence of the $\chi ^{\prime
\prime }$-peaks. The huge effect in $U$(0, 0) is due to a strong weakening
in the connectivity of the intergranular network as a consequence of the Ru
magnetization in the grains. The impact on $J_{C}$(0, 0)$_{Ru}$ is less
intense due to the small average grain size.

The value of $J_{C}$(0, 0)$_{Ru}$ can be calculated by an alternative
approach, still within the framework of the critical state model. Equation
(1) can be written as

\begin{equation}
dH(x)/dx=J_{C}(0,T)\frac{H_{0}/2}{\left\vert H(x)\right\vert +H_{0}/2}+\frac{%
kT}{2R_{g}^{2}\Phi _{0}}\ln \frac{f}{f_{0}}
\end{equation}%
\newline
where the dependence of $J_{C}$ on the local magnetic field $H(x)$ is
modeled by a Kim-type envelope of the Fraunhofer patterns associated to the
distribution of junction qualities in the polycrystal. Here, $H_{0}$=$\Phi
_{0}$/$\mu _{0}A_{J}$, where $\mu _{0}$ is the permeability of free space, $%
A_{J}$=2$R_{g}$[2$\lambda _{L}$($T$)] is the field penetrated junction area
and $\lambda _{L}$($T$) is the intragrain London penetration length.
Evaluating Eq. (5) for the full penetration condition of the bar by an
external field at $T$ = 0 and integrating we obtain

\begin{equation}
-\frac{H^{\ast 2}(0)}{2H_{0}(0)}-H^{\ast }(0)+J_{C}(0,0)_{Ru}\left( \frac{a}{%
2}\right) =0
\end{equation}%
\newline
where $a$ is the thickness of the bar.

The use of Eq. (6) for the calculation of $J_{C}$(0, 0)$_{Ru}$ requires the
knowledge of $\lambda _{L}$(0). However, it must be kept in mind that the
intragrain London penetration length has some peculiarities in the
ruthenate-cuprates. It has been demonstrated for Ru-1222 that the intragrain
superfluid density 1/$\lambda _{L}{}^{2}$ does not follow by far the linear
correlation with $T_{SC}$ observed for homogeneous cuprates.\cite{Xue02}
Also, it was found that $\lambda _{L}$ is very sensitive to the partial
pressure of oxygen during the final annealing, varying between 0.4 $\mu $m ($%
T_{SC}$ = 40 K) and 1.8 $\mu $m ($T_{SC}$\ = 17 K).\cite{Xue02} For Ru-1212,
samples with intragrain transition temperatures higher than 20 K show $%
\lambda _{L}$ at 5 K as large as 2-3 $\mu $m.\cite{Xue01} Thus, there is
considerable uncertainty in the choice of the appropriate value of $\lambda
_{L}$ to determine $J_{C}$(0, 0)$_{Ru}$ in our sample using Eq.(6). Instead,
we look for a confirmation of the validity of the critical state model by
taking the $J_{C}$(0, 0)$_{Ru}$ value obtained from the frequency shift, and
derive $\lambda _{L}$(0). Taking $J_{C}$(0, 0)$_{Ru}$=110 A/cm$^{2}$, $R_{g}$
= 1 $\mu $m, $H$*(0) = 5 Oe, and $a\ $= 1.7 mm we obtained $\lambda _{L}$(0)
= 2.2 $\mu $m, in good agreement with the value $\lambda _{L}$(5 K) = 1.8 $%
\mu $m reported for Ru-1222 ceramic with an intragrain transition
temperature $T_{SC}$ = 17 K (near to our value of $T_{SC}$ = 22 K) through
the particle size dependence of the real part of the susceptibility.\cite%
{Xue02} We remark that $\lambda _{L}$(0) is larger in comparison
to YBCO and other optimally-doped high T$_{c}$ cuprates, and
comparable to the average grain size of Ru-1222. This should be
related not only to the underdoped character of the
ruthenate-cuprates, as revealed by Hall effect and thermopower
measurements,\cite{Tallon,Croone} but also to a very distinctive
feature of these compounds. Both Ru-1212 and Ru-1222 exhibit
structural domains of coherent rotated RuO$_{6}$ octahedra,
separated by antiphase boundaries with local distortions and
defects.\cite{McLaughlin} This characteristic has been proposed to
be the source of the strong intragrain granularity observed in
these compounds, with the boundaries acting as
intragrain Josephson junctions between the structural domains.\cite%
{Garcia03a} Within this scenario, the magnetic field penetrates the grains
not only through their crystallographic borders but also through the
antiphase boundaries, increasing the penetrated volume and leading to an
enhanced effective $\lambda _{L}$ and to a small and sometimes missing
Meissner signal. Therefore, the obtained $\lambda _{L}$(0) provides a
reliable estimate, and confirms the validity of the critical state approach.

Finally, we address the magnitude of the local field at the intergranular
junctions due to the Ru magnetization, $H_{Ru}$. The use of the critical
state model, leading to a decreasing field profile in the ceramic as its
center is approached, is valid only if the external field is comparable to $%
H_{Ru}$. In the case where $h_{ac}$ is much smaller than $H_{Ru}$, its
effect on the connectivity of the weak link network is negligible, and no
amplitude dependence for the position of the $\chi ^{\prime \prime }$-peaks
would be observed. In other words, a large $H_{Ru}$ will result in a flat
intrinsic field profile unaffected by $h_{ac}$. Magnetic fields of the order
of 600-700 Oe have been measured by muon spin rotation\cite{Bernhard} and
Gd-electron paramagnetic resonance\cite{Fainstein} in sites located near to
the RuO$_{2}$ layers. However, the dipolar field rapidly decays with
distance, and the local field at the junctions can be considerably smaller.
Defects and imperfections in the region of the intergrain boundaries can
locally affect the magnetic order of the Ru moments, diminishing the
effective field. An estimation of $H_{Ru}$ can be performed using the value
reported for the intergrain critical current in the isomorphous
non-ferromagnetic Nb-1222 compound ($J_{C\text{~}Nb}$ = 1545 A/cm$^{2}$ at $%
T $ = 5 K) \cite{Felner03}). Assuming similar superconducting properties,
and keeping the Kim-type model to account for the dependence of $J_{C}$ with
the local magnetic field we can write%

\begin{align}
J_{C}(H_{ext},T=0)_{Ru}
& =J_C(H_{ext}=0,T=5K)_{Nb}\times\nonumber\\
& \,\,\,\,\,\times
\biggl[\frac{H_{0}/2}{|H(x)+H_{Ru}|\,+\,(H_0/2)}\biggr]
\end{align}
,where $H(x)$ and $H_{Ru}$ must be added taking into account their
relative
orientations. Evaluating Eq. 7 for $H_{ext}=$ $h_{ac}$ = 0 and taking $%
\lambda _{L}$(0) = 2.2 $\mu $m, $R_{g}$ = 1 $\mu $m [giving $H_{0}$(0) = 2.3
Oe] and $J_{C}$(0, 0)$_{Ru}$ = 110 A/cm$^{2}$, we obtain $H_{Ru}\cong $15
Oe. An intrinsic local field at the intergranular junctions of this
magnitude is large enough to greatly affect the connectivity of the network,
but still leaving it sensitive to the action of external oscillating fields
of a few Oersted.

In summary, a detailed study of $\chi ^{\prime \prime
}$($h_{ac},f,T$) curves in polycrystalline Ru-1222 allowed for the
first time to our knowledge a quantitative characterization of the
intergrain coupling in a ruthenate-cuprate. The intergranular
pinning potential and the critical current density at zero field
and temperature were determined. A critical state model, including
flux creep effects, was found to properly describe the $\chi
^{\prime \prime }$ behavior. The pinning potential showed a very
strong decrease in comparison to the high T$_{c}$ cuprates. This
is ascribed to the effects of the Ru magnetization in the grains
on the connectivity of the weak link network, giving a local field
at the junctions of about 15 Oe. The approach yields a large value
for the intragrain London penetration length, which matches with
the strong intragrain granularity effects observed in Ru-1222.

\begin{center} {\bf{Acknowledgements}}
\end{center}

This work was partially supported by CNPq. S.G. was financed by
FAPERJ.

\end{document}